# Ultraviolet Spectra of Local Galaxies and their Link with the High-$z$ Population

Claus Leitherer

*Space Telescope Science Institute, 3700 San Martin Dr., Baltimore, MD 21218, USA*

**Abstract.** The new generation of 8 to 10m class telescope is providing us with high-quality spectral information on the rest-frame ultraviolet region of star-forming galaxies at cosmological distances. The data can be used to address questions such as, e.g., the star-formation histories, the stellar initial mass function, the dust properties, and the energetics and chemistry of the interstellar medium. We can tackle these issues from a different angle by comparing the spectral properties of high-redshift galaxies to those of their counterparts in the local universe. I give a review of recent developments related to observations and empirical modeling of the ultraviolet spectra of local galaxies with recent star formation. The emphasis is on the youngest stellar populations with ages less than 100 Myr. Current uncertainties will be discussed, and areas where progress is needed in the future are highlighted.

## INTRODUCTION

The first scientifically useful ultraviolet (UV) spectra of astronomical objects outside the solar system were obtained in the 1960s when the 3-axis star-pointing stabilization system of the Aerobee sounding rockets permitted the acquisition of sufficiently deep spectrograms [1]. However, significant numbers of UV spectra of normal galaxies could not be accumulated until the advent of the IUE satellite, which had the capability of collecting multi-hour exposures necessary for extragalactic studies [2]. The first- (FOS), second- (GHRS), and third-generation (STIS) spectrographs of HST, together with HUT, each led to order-of-magnitude improvements of spectral resolutions and progressively higher signal-to-noise. In parallel with the progress in satellite-UV astronomy, a new generation of 8 to 10-m class ground-based telescopes went on-line during the past decade. The telescopes have produced restframe-UV spectra of star-forming galaxies at cosmological distances whose quality rivals and often exceeds that of their local counterparts [3]. A comparison of the average spectra of 16 local star-forming galaxies and of those of 811 Lyman-break galaxies (LBG) suggests striking similarity [4].

In this review, I will highlight the overall spectral similarity between local and distant star-forming galaxies but at the same time point out some subtle but significant differences. After a brief summary of the basic galaxy properties, I will cover the stellar populations, the neutral and ionized interstellar medium (ISM), Lyman-α, and the escape of Lyman continuum radiation.

# GALAXY PROPERTIES

Local star-forming galaxies targeted for UV spectroscopy are necessarily UV-bright, a bias imposed by the low quantum efficiency of UV detectors and the relatively small telescope sizes. Morphologically, these galaxies tend to be of late Hubble types, and they include blue compact galaxies, H II galaxies, and nuclear starbursts [4]. Stellar masses are of order $10^9$ M$_\odot$, and absolute magnitudes are in the range $-16 < M_B < -19$. Oxygen abundances are as low as 1/20$^{\text{th}}$ Z$_\odot$ and as high as Z$_\odot$, with typical values similar to those of the Magellanic Clouds. Overall, local UV-bright galaxies cover a parameter space that is similar to that occupied Lyman-α emitters at high redshift [5], but with the important difference of generally weak Lyman-α emission. The local sample is often quoted as the counterpart of LBGs. While the two samples are similar in many respects, it is important to realize that the average luminosities and masses of LBGs are 1 to 2 orders of magnitude higher than for the local sample.

Since these galaxies were selected based on their UV brightness, they tend to have low dust reddening. As a result, their morphologies are often quite similar at different wavelengths, in particular when going from the UV to the optical. A comparison of GALEX far- and near-UV and SDSS optical imagery supports this view [6]. Nevertheless, examples of UV-bright star-forming galaxies with strong local dust attenuation exist. The UV may very well provide a rather biased view of the actual star-formation activity. A striking example is the interacting galaxy pair VV 114 [7], whose two members show a strong color contrast. One component is dominated by a blue, high surface brightness complex of regions with a relatively weak near-infrared (IR) nucleus. The other component is much redder and brighter in the near-IR but inconspicuous in the UV. If this system were observed at high redshift in the absence of spatial information, the apparently coincident UV and IR would arise in spatially disjoint regions, and correlating them would be meaningless.

# STELLAR POPULATION

The satellite-UV traces the most recently formed stars with masses of ~5 M$_\odot$ and above. The continuum below the Balmer break comes from late-O and early-B stars. Superimposed on the continuum are strong, broad, blueshifted absorption lines, sometimes with emission components, from O stars of all spectral type. On average, these O stars have masses of order 50 M$_\odot$. The most prominent spectral lines are O VI λ1035, N V λ1240, Si IV λ1400, and C IV λ1550 [8].

The UV lines originate in powerful stellar winds with stellar-mass-dependent properties and cover a wide range of ionization potentials from a few eV to 114 eV (O VI). This makes them suitable for studying the mass distribution and eventually the initial mass function (IMF) of the most massive stars in the mass range between 10 and 100 M$_\odot$. A major outcome of numerous spectroscopic UV studies of local star-forming galaxies is the ubiquity of a single Salpeter-like IMF in this mass range. This result has independently been confirmed by other methods, such as the photo-ionization modeling of optical nebular emission lines [9].

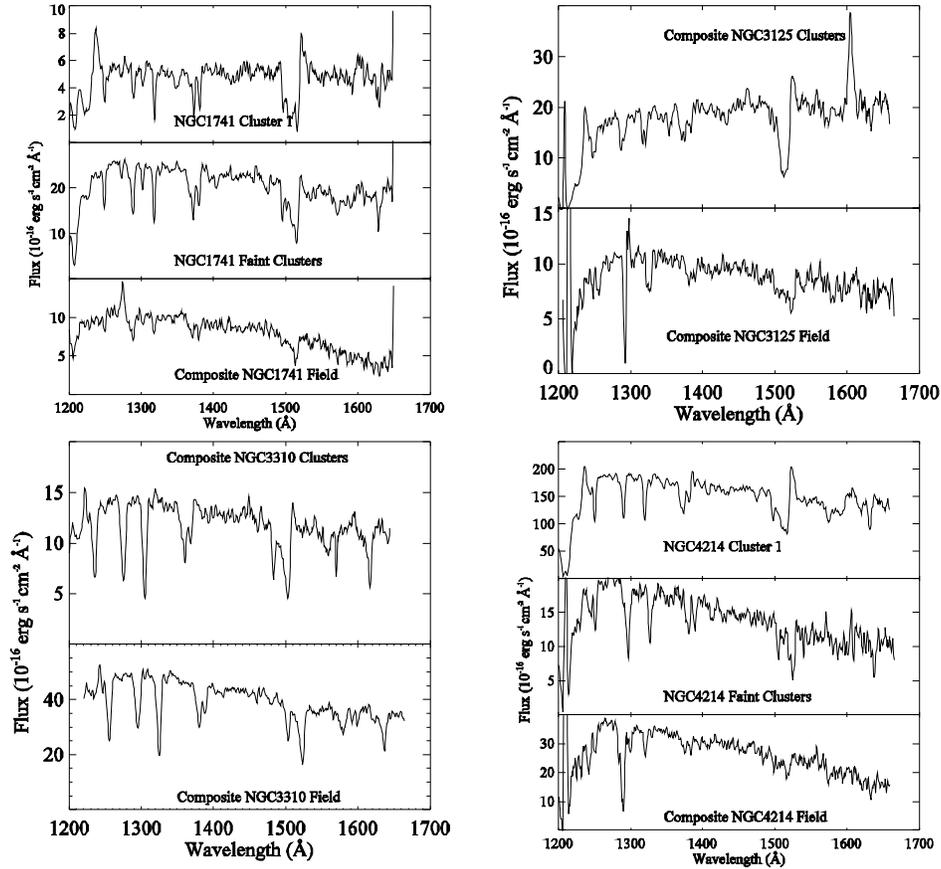

**FIGURE 1.** Composite cluster and field spectra for four dwarf galaxies obtained with STIS. The field spectra have weaker N V, Si IV, and CIV lines indicating a deficit of the most massive stars. From [10].

However, one should be aware that essentially all these data were taken through narrow slits extending over length scales of order 10 pc in each galaxy. The apertures typically encompass one or more bright star clusters, which constitute the local peak of the UV light in each galaxy. Fig. 1 illustrates this point. In this case, long-slit spectroscopy with HST's STIS allows separate studies of both the clusters and the intercluster light, which is the diffuse, unresolved stellar emission. Comparison of the cluster and intercluster light suggests weaker stellar NV, Si IV, and C IV lines in the field. This translates into a deficit of very massive field stars. One interpretation could be a steeper field IMF, which has fewer O stars. Alternatively, an age effect could be responsible: field stars are older on average because field stars are the relics of dissolved star clusters whose lifetimes of order 10 Myr are longer than O-star lifetimes. Therefore, massive O stars may disappear before cluster dissolution and never contribute to the field population [11].

The lesson learned for the interpretation of the restframe UV spectra of high-$z$ galaxies is to be aware of the bias that is inherent in local spectra. The latter usually refer to only a few bright star clusters whose light contribution to the total is only a few percent and whose stellar population may not be representative for the galaxy as a whole. In contrast, spectra of distant galaxies encompass a much larger volume, and using local template spectra may introduce a significant bias.

## NEUTRAL AND IONIZED ISM

In addition to the already discussed stellar-wind lines, numerous strong interstellar absorption lines are located in the wavelength region below 3000 Å. The lines can easily be distinguished from the stellar lines by their line widths, whose values of a few $10^2$ km s$^{-1}$ are smaller by almost an order of magnitude.

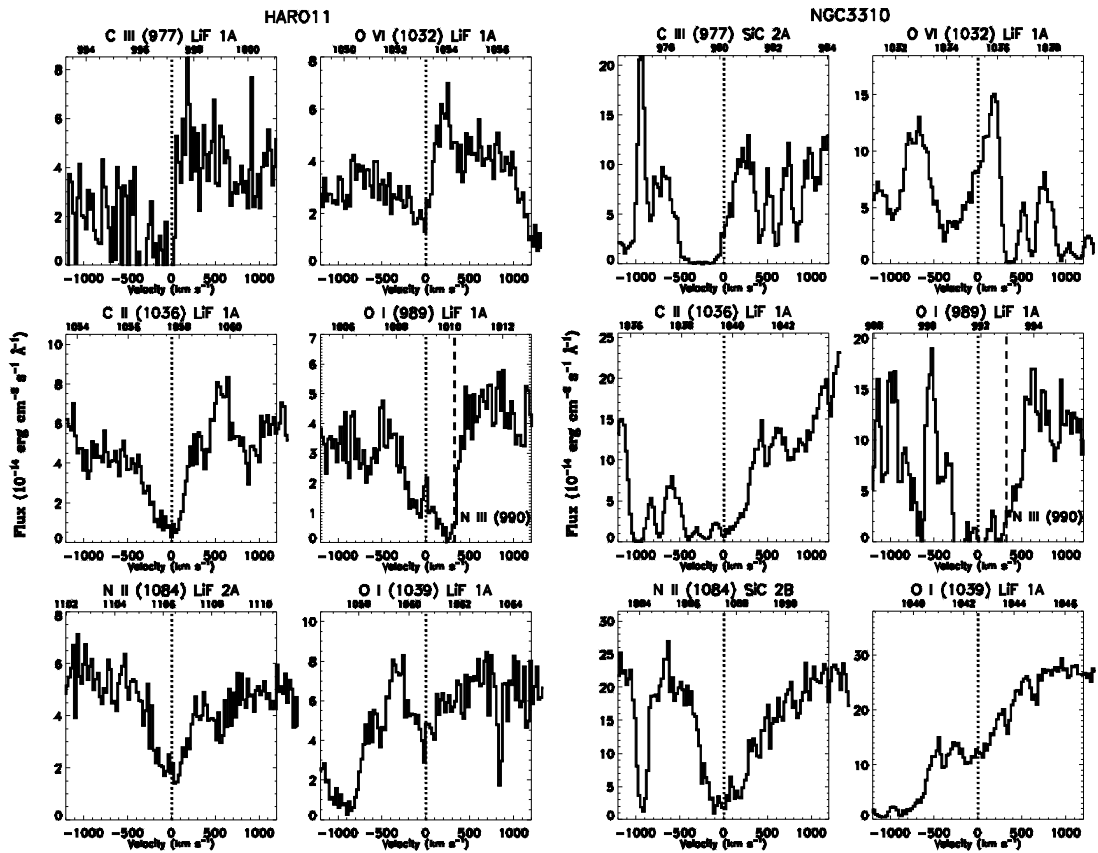

**FIGURE 2.** UV absorption lines observed with FUSE. Left: Haro 11; right: NGC 3310. Note: O I λ989 is blended with N III λ990. From [12].

The FUSE satellite was specifically optimized for observations of the Galactic and extragalactic ISM. Its wavelength coverage was from 1175 Å down to the Lyman limit at 912 Å, at a resolving power of 20,000. The ISM lines in a sample of 16 star-forming galaxies observed with FUSE by [12] are all blueshifted and asymmetric. In Fig. 2 I have reproduced the data for two of their program galaxies. Velocity displacements of hundreds of km s$^{-1}$ can be seen. These displacements are indicative of galaxy-wide outflows, also known as galactic superwinds [13]. The existence of such outflows in essentially all local star-forming galaxies has been demonstrated from observations of the cool (Na I), warm (Hα), and hot (O VI and X-rays) gas. Stellar winds and supernovae support a pressure-driven outflow, which expands along the direction of the maximum density gradient.

The FUSE data of [12] suggest a trend of larger outflow velocities in galaxies with larger specific star-formation rate *SFR/M*, where *SFR* is measured from the combined UV and IR luminosities, and the *K*-band luminosity is used as a proxy for stellar mass. Since the specific star-formation rate increases with redshift [14], one may speculate that galactic superwinds are even more pronounced and prevalent in, e.g., LBGs. If so, the importance of the escape of processed matter from galaxies into the surrounding intergalactic medium (IGM) and leakage of ionizing radiation will increase from low to high redshift.

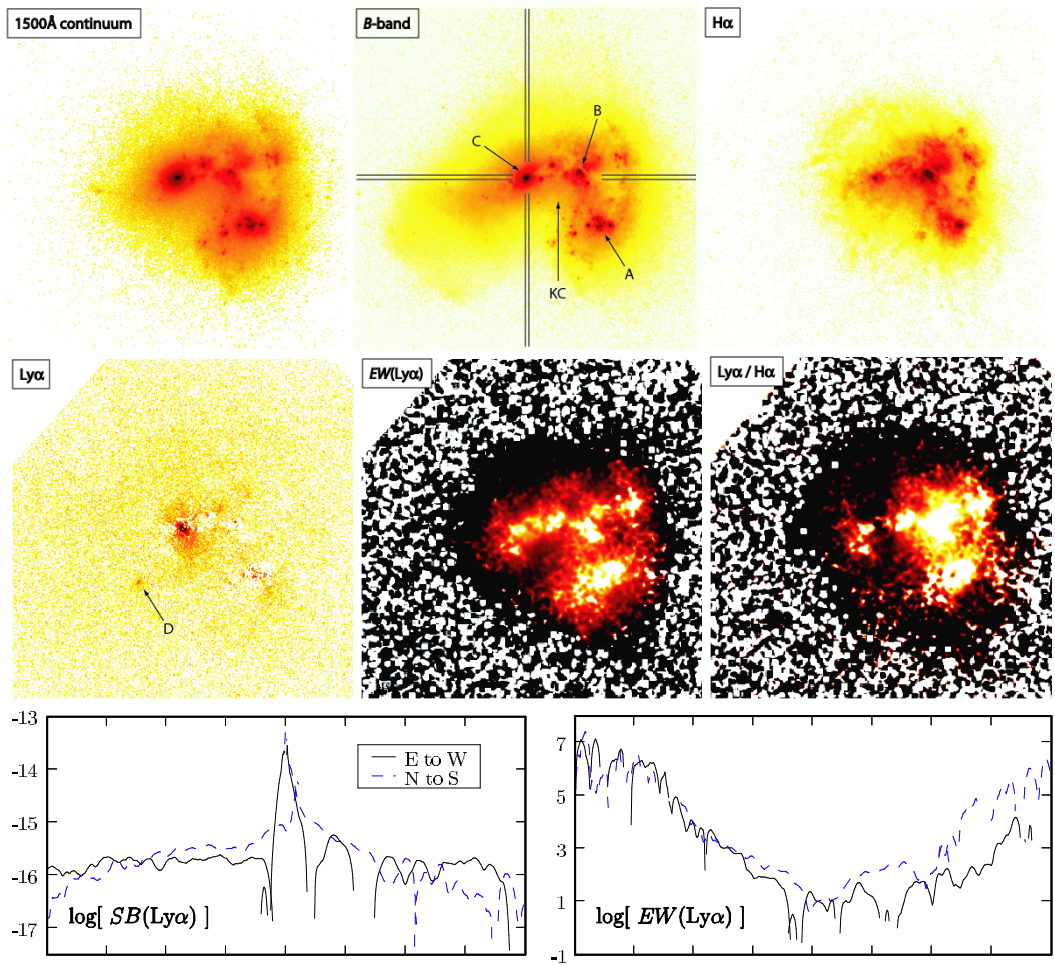

**FIGURE 3.** Haro 11 as seen by HST. Field sizes are $20 \times 20$ arcsec$^2$, corresponding to $8.1 \times 8.1$ kpc$^2$. The individual images show Haro 11 in the 1500 Å continuum, the B-band, Hα, Lyman-α, the Lyman-α equivalent width, and the Lyman-α/Hα ratio. The bottom panels show spatial cuts along rows and columns for the Lyman-α flux (left) and equivalent width (right). From [18].

# LYMAN-ALPHA

If the nebular Lyman-α line behaved like an ideal recombination line, its predicted equivalent width (*EW*) for a standard young stellar population is of order $10^2$ Å [15]. Such large *EW* values are never observed in local star-forming galaxies, which often display Lyman-α as a damped absorption profile [16]. This is surprising, as Lyman-α is typically observed as a strong emission line in high-redshift star-forming galaxies, whose properties are otherwise quite similar to their low-redshift counterparts [17].

A major limitation of current UV spectrographs is their lack of spatial resolution and/or small spatial coverage. Narrow-band Lyman-α imaging can provide invaluable complementary information. Fig. 3 summarizes the results of HST/ACS imagery of Haro 11, whose luminosity ($M_B = -20.5$) and oxygen abundance (log O/H +12 = 7.9) make it an excellent analog of an LBG [18]. Most Lyman-α photons are emitted in the nucleus (bottom left) but since the stellar UV continuum is even more peaked towards the nucleus (top left), the Lyman-α equivalent width is very small in the center (bottom right). The equivalent width is in fact quite high outside the nucleus where the neutral hydrogen column is low but the absolute number of ionizing photons produced there is too small to be of importance. Only 3% of the Lyman-α photons expected to be observed based on the Hα recombination flux escape.

The reason for the higher Lyman-α escape fraction in galaxies at high redshift may lie with the ISM porosity and dynamics. Galactic winds are more powerful at high redshift, leading to increased stirring of the ISM and creating an effective Lyman-α escape mechanism.

# LYMAN CONTINUUM

A standard star-forming population emits approximately 10% of its luminosity as ionizing radiation below 912 Å [19]. Most of this radiation is absorbed by the ambient neutral hydrogen and by dust, as suggested by the observed recombination lines. Yet the possibility exists, and even seems likely, that some fraction of the ionizing photons will escape from both the H II regions and the diffuse ISM. If so, star-forming galaxies could be an important source for the cosmological ionizing background radiation.

One can measure the escape fraction either in local galaxies using a far-UV detector or in galaxies at cosmological redshift, whose restframe UV then becomes accessible from the ground with 8-m class telescopes. Either technique has its advantages and disadvantages. The "local" approach faces the obvious challenge of extreme UV observations, whereas the "cosmological" measurement must account for the radiative transfer in the IGM.

The FUSE survey of [12] can shed additional light on this issue. Nine program galaxies have sufficiently high velocity to shift the intrinsic Lyman continuum out of the Galactic foreground H I absorption. The spectra of five of these galaxies are plotted in Fig. 4. No significant Lyman continuum emission is detected in any of the target galaxies. The upper limits on the fluxes, when combined with simple models for the geometry of the ISM, permit relatively stringent constraints on the Lyman continuum escape fractions. For a picket-fence model of the ISM, average escape fractions of less than about 1% are found.

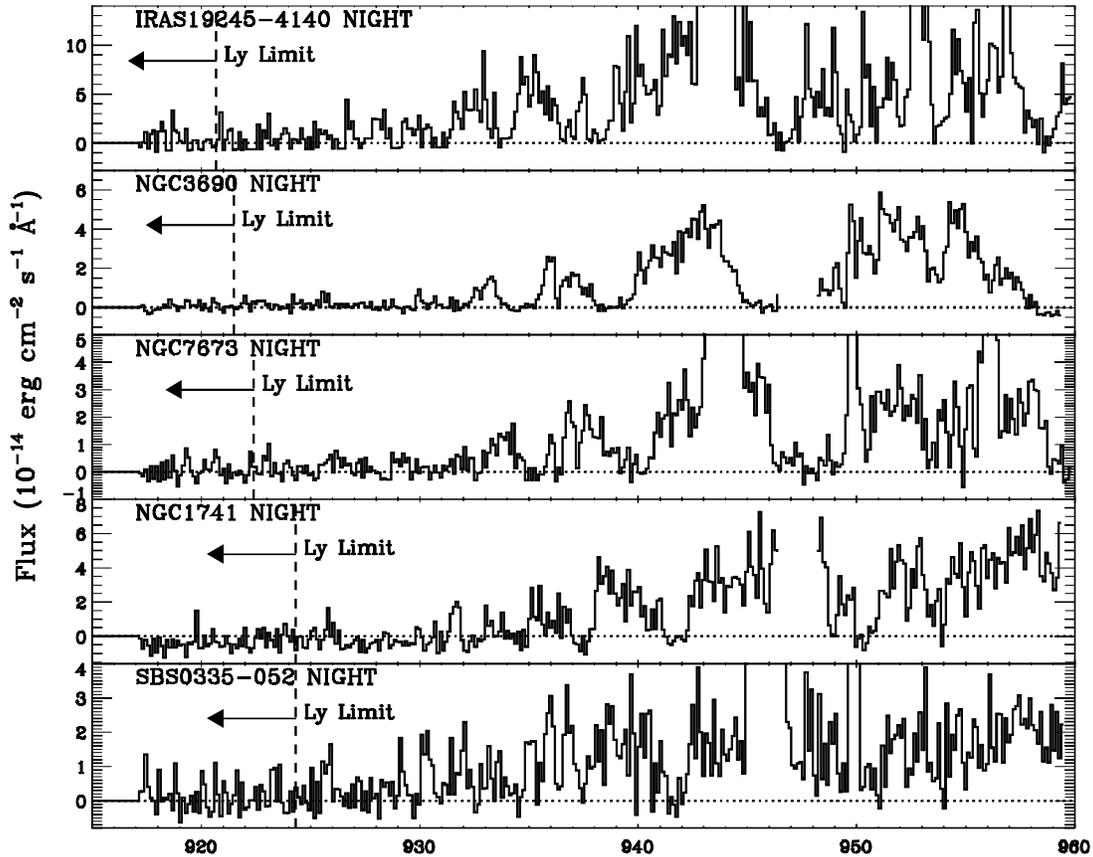

**FIGURE 4.** Lyman continuum regions of five galaxies in order of increasing redshift observed with FUSE. The wavelengths are in the observed frame. The intrinsic Lyman continuum is to the left of each vertical dashed line. There is no convincing evidence of Lyman continuum emission in these spectra. From [12].

Alternatively, the strength of the interstellar absorption lines can be used to infer the neutral hydrogen opacity, and therefore the escape probability of hydrogen ionizing photons. The FUSE spectral range includes species of several abundant elements with ionization stages close to dominant, thus minimizing uncertain model assumptions. The interstellar lines imply hydrogen column densities which limit the Lyman photon escape to less than a few percent, consistent with the direct measurement below 912 Å.

The FUSE result mirrors previous studies of the Lyman continuum in local star-forming galaxies, which uniformly failed to detect significant Lyman continuum radiation [20]. The evidence at high redshift is less clear. [21] reported a detection in the composite spectrum of 29 LBGs with average redshift $z = 3.40 \pm 0.09$. On the other hand, a different LBG sample studied by [22] implied an upper limit 4.5 times lower than inferred from the composite spectrum of [21]. Strikingly, two out of the 14 sample galaxies observed by [21] show a clear detection of Lyman continuum radiation whereas the remaining twelve are non-detections. Averaging over the whole sample of [22] leads to a mean escape fraction of 14%, with a large variation from galaxy to galaxy.

The empirical result of a higher Lyman continuum escape fraction from lower to higher redshift may again be understood in terms of more violent star formation in the early universe. More powerful galactic superwinds that are initiated and supported by stellar winds and supernovae increase the ISM porosity and create escape paths for the stellar ionizing radiation. Better observational statistics and quantitative modeling are required for determining how the escaping radiation compares to the contribution of optically selected quasars at the same redshift and whether star-forming galaxies are ultimately responsible for the reionization of the early universe.